\documentclass[aps,prl,reprint,superscriptaddress]{revtex4-1}
\newcommand{\figSize}{0.5}

\usepackage{graphicx}
\usepackage{epstopdf}
\newcommand{\ExState}{|\mathrm{E}_\mathrm{x}\rangle}
\newcommand{\EyState}{|\mathrm{E}_\mathrm{y}\rangle}
\newcommand{\Ex}{|0\rangle\leftrightarrow|\mathrm{E}_\mathrm{x}\rangle}
\newcommand{\Ey}{|0\rangle\leftrightarrow|\mathrm{E}_\mathrm{y}\rangle}
\newcommand{\gPar}{g^{(2)}_{||}}
\newcommand{\gPerp}{g^{(2)}_{\bot}}
\newcommand{\gPSB}{g^{(2)}_{\mathrm{PSB}}}
\newcommand{\Vapp}{V_\mathrm{app}}

\begin{document}

\title{Quantum interference of single photons from remote nitrogen-vacancy centers in diamond}

\author{A. Sipahigil}
\affiliation{Department of Physics, Harvard University, Cambridge, Massachusetts 02138, USA}

\author{M. L. Goldman}
\affiliation{Department of Physics, Harvard University, Cambridge, Massachusetts 02138, USA}

\author{E. Togan}
\affiliation{Department of Physics, Harvard University, Cambridge, Massachusetts 02138, USA}

\author{Y. Chu}
\affiliation{Department of Physics, Harvard University, Cambridge, Massachusetts 02138, USA}

\author{M. Markham}
\affiliation{Element Six Ltd, Kings Ride Park, Ascot SL5 8BP, UK}

\author{D. J. Twitchen}
\affiliation{Element Six Ltd, Kings Ride Park, Ascot SL5 8BP, UK}

\author{A. S. Zibrov}
\affiliation{Department of Physics, Harvard University, Cambridge, Massachusetts 02138, USA}

\author{A. Kubanek}
\email[]{kubanek@fas.harvard.edu}
\affiliation{Department of Physics, Harvard University, Cambridge, Massachusetts 02138, USA}

\author{M. D. Lukin}
\affiliation{Department of Physics, Harvard University, Cambridge, Massachusetts 02138, USA}



\begin{abstract}
We demonstrate quantum interference between indistinguishable photons emitted by two nitrogen-vacancy (NV) centers in distinct diamond samples separated by two meters. Macroscopic solid immersion lenses are used to enhance photon collection efficiency. Quantum interference is verified by measuring a value of the second-order cross-correlation function $g^{(2)}(0) = 0.35 \pm 0.04<0.5$. In addition, optical transition frequencies of two separated NV centers are tuned into resonance with each other by applying external electric fields. Extension of the present approach to generate entanglement of remote solid-state qubits is discussed.
\end{abstract}


\maketitle

The interference of two identical photons impinging on a beamsplitter leads to perfect coalescence where both photons leave through the same output port. This fundamental effect, known as Hong-Ou-Mandel (HOM) interference \cite{Hong1987}, is a consequence of bosonic statistics for indistinguishable particles. HOM interference has been demonstrated using single photon pairs from parametric down-conversion \cite{Ou2007} and delayed photons from a single photon source \cite{Santori2002,Legero2004,Kiraz2005}. HOM interference has recently drawn attention as a resource for entanglement generation between distinct single-photon emitters with many potential applications in quantum information science \cite{Duan2010}. The effect has been observed for photons emitted by pairs of atoms \cite{Beugnon2006} and trapped ions \cite{Maunz2007}, and has been used for entanglement generation of remote trapped ions \cite{Moehring2007}. While isolated atoms and ions, which are nominally identical, are a natural source of indistinguishable photons, extending these ideas to condensed matter systems can be challenging since two solid-state emitters are generally distinguishable because of their different local environments. This Letter demonstrates quantum interference of two photons produced by nitrogen-vacancy (NV) impurities in distinct diamond samples separated by two meters. Complementing the recent work involving other solid-state systems \cite{Bernien2012,Lettow2010,Flagg2010,Patel2010}, the present solid-state realization is particularly significant, since electronic and nuclear spins associated with NV centers can be used as a robust solid-state qubit memory, yielding potential scalable architectures for quantum networks \cite{Childress2006,Dutt2007}. Specifically, in combination with a recent demonstration of spin-photon entanglement \cite{Togan2010}, the present work paves the way for entanglement generation between remote solid-state qubits.

\begin{figure}[h!]
\begin{center}\includegraphics[width = \figSize \textwidth]{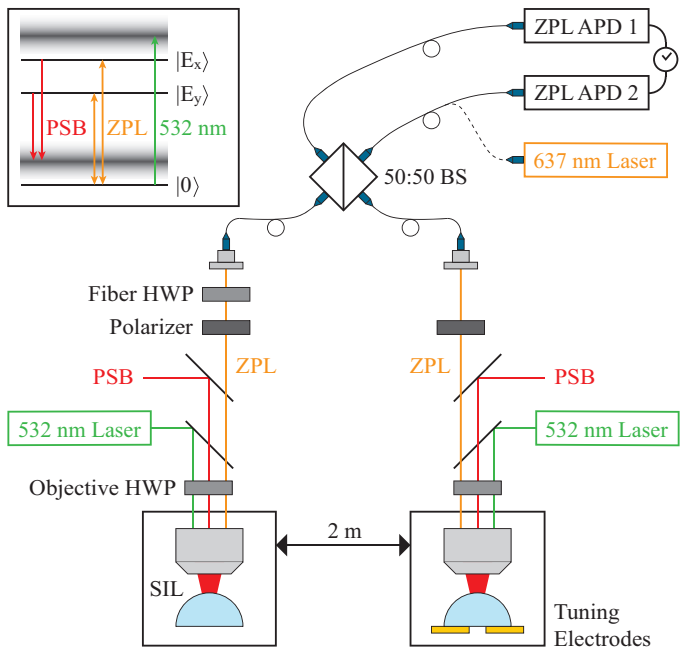}
\caption[Schematic of apparatus]{Schematic of the apparatus.  Two diamond SILs containing multiple NV centers are housed in continuous helium flow cryostats 2 m apart.  Each SIL is addressed by a separate confocal microscopy setup, which includes a path for excitation at 532 nm, a collection path for the phonon sideband (PSB), and a collection path for the zero-phonon line (ZPL).  The three optical paths are superimposed using dichroic mirrors. The ZPL collection path passes through a half-wave plate (HWP) and an additional band-pass filter (633-647 nm) before being coupled into a polarization-maintaining single-mode 50:50 fiber beamsplitter.  The two beamsplitter output arms are connected to a pair of avalanche photodiodes (APDs), completing the Hanbury Brown and Twiss detection setup.  An excitation laser at 637 nm is connected in place of ZPL APD 2 to acquire the photoluminescence excitation spectra.  Electrodes for electric field tuning are installed in one cryostat.  The inset shows a simplified level structure, including resonant excitation and emission into the ZPL at 637 nm, red-shifted emission into the ground state PSB, and non-resonant excitation into the excited state PSB at 532 nm.}
\label{fig.Apparatus}
\end{center}
\end{figure}

Unlike those associated with atoms in free space, the optical properties of NV centers embedded in a solid state vary substantially from emitter to emitter, especially in distinct samples. This inhomogeneity is due to variation in the local environments of NV centers and, in particular, to variation in the local strain. Furthermore, coincidence experiments are limited by the collection efficiency for light emitted by the NV center.  While a wide variety of approaches are currently being explored to enhance the collection efficiency  \cite{Babinec2010,Faraon2011,Englund2010,Kolesov2009}, we here utilize solid immersion lenses (SILs) fabricated from bulk diamond \cite{Siyushev2010} to facilitate the efficient collection of narrowband photons with identical properties from distant diamond samples. The SILs improve the collection efficiency by minimizing total internal reflection at the air-diamond interface, which is significant because of the high refractive index ($n_d=2.4$) of the diamond host. We measure enhancement factors in the range of 6-10, depending on the position of the NV center inside the SIL. Very recently, microfabricated SILs have been used to observe HOM interference from two NV centers separated by roughly 20 $\mu$m on the same diamond chip \cite{Bernien2012}.

In our experiment, we use two 1.0-mm diameter SILs that are fabricated from bulk electronic grade diamond and cut along the (100) crystal plane. The SILs are kept at a temperature of 8K in continuous flow helium cryostats that are separated by two meters, as shown in Fig. \ref{fig.Apparatus}. We characterized the spectral properties of several NV centers using photoluminescence excitation (PLE) spectroscopy. This technique involves collecting the red-shifted PSB emission while an external-cavity diode laser is scanned across the ZPL transitions (see the inset of Fig. \ref{fig.Apparatus}). The resulting PLE spectra reveal FWHM linewidths in the range of 50-250 MHz for individual transitions of NV centers in both SILs. These linewidths, although broadened by charge fluctuations in the vicinity of the NV center due to the ionization of nearby charge traps \cite{Tamarat2006}, are comparable to the narrowest linewidths observed in both synthetic and natural bulk diamond samples \cite{Fu09,Tamarat2006}.

To obtain identical photons from two NV centers, the NV centers need to have transitions that are spectrally overlapping, and the emission from these individual transitions for each NV needs to be isolated. By performing simultaneous PLE scans on NV centers in the two SILs with a single laser, we can directly measure the relative detuning of their optical transitions. In our experimental sequence, a 5 $\mu$s pulse of green light initializes the NV center into the electronic spin sublevel of the triplet ground state with $m_s = 0$ ($|0\rangle$) \cite{Manson2011}. Therefore, we only collect fluorescence from the NV center when the laser is resonant with transitions from the $|0\rangle$ state to the $\ExState$ or $\EyState$ states, as shown in inset of Fig. \ref{fig.Apparatus} \cite{Batalov2009}. We need to select a pair of NV centers such that one transition in the first NV center is resonant, or can be tuned into resonance, with one transition in the second.

\begin{figure}
\begin{center}\includegraphics[width = \figSize \textwidth]{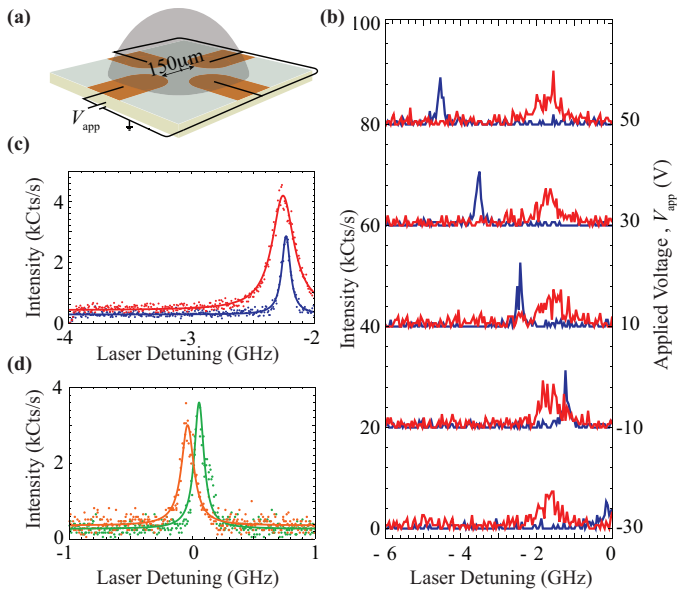}
\caption[Electric Field Tuning]{Electric field tuning of optical transitions. (a) Four Cr/Au gates deposited on silicon. The central gap has a diameter of 150 $\mu$m. In this experiment, only one of the gate voltages was swept while the others were kept grounded. (b) PLE spectra for different applied gate voltages. The gate voltage $\Vapp$ is varied from -30 to 50 V for NV1, which creates an electric field of up to 0.5 MV/m at NV1.  The  $\Ex$ transition of NV1 (blue) is tuned across the $\Ex$ transition of NV2 (red). For different $\Vapp$, PLE spectra are offset by 20 kCts/s for clarity. (c) Linewidth measurement under electric field tuning. On resonance, the measured FWHM linewidths are $85\pm2$ MHz for NV1 (blue) and $217\pm4$ MHz for NV2 (red). The detuning of the optical transitions in two samples is $25\pm2$ MHz. (d) Linewidth measurement for the NV centers used for the HOM measurement. The measured linewidths are $88\pm3$ MHz (green) and $106\pm4$ MHz (orange), and the detuning is $93\pm15$ MHz without electric field tuning.}
\label{fig.Tuning}
\end{center}
\end{figure}

We next demonstrate control over the optical properties of NV pairs to compensate for strain-induced spectral inhomogeneities.  We make use of the DC Stark effect to actively minimize the detuning between the selected transitions \cite{Tamarat2006,Bassett2011}. Electric fields perpendicular to the NV axis vary the splitting between $\ExState$ and $\EyState$ states and parallel fields shift both transitions together.
This allows complete control over the optical transition frequencies \cite{Maze2011,Bassett2011}. In order to apply the desired electric field, we place one of the SILs on top of a silicon wafer deposited with four electrodes, comprised of 40 nm Au on a Cr adhesion layer. The gate geometry is shown in Fig. \ref{fig.Tuning}(a). We apply a bias voltage, $\Vapp$, on one of the electrodes while keeping the other three grounded.  In Fig. \ref{fig.Tuning}(b), the $\Ex$ transition of NV1 (blue) is tuned across the $\Ex$ transition of NV2 (red) by varying the applied voltage $\Vapp$ from -30 V to 50 V,  which creates an electric field of up to 0.5 MV/m at NV1. At $\Vapp=-2.9$ V, shown in Fig. \ref{fig.Tuning}(c), the detuning between the two transitions is reduced to $25\pm2$ MHz from an initial value of $270\pm15$ MHz. We measure linewidths of  $85\pm2$ MHz for NV1, which has been tuned, and $217\pm4$ MHz for NV2, which has not been tuned. Similarly, we do not observe a systematic change of the linewidths with applied external fields in several other NV centers.


For the HOM interference measurement, we excite the NV centers with green light and collect the ZPL emission. Because the green excitation ionizes charge traps in the diamond lattice \cite{Bassett2011} and these charge dynamics can change the total electric field at the NV center, the time during which we can collect fluorescence at the tuned frequency is significantly reduced. For this reason, we select NV centers for the HOM measurement whose transitions, shown in Fig. \ref{fig.Tuning}(d), are inherently detuned by $93\pm15$ MHz and have linewidths of $88\pm3$ MHz and $106\pm4$ MHz, eliminating the need for electric field tuning.

We want to isolate the emission from the selected transitions so that we only collect spectrally overlapping ZPL photons for the HOM measurement. ZPL emission is separated from the PSB emission using a dichroic mirror and a spectral band-pass filter, as described in the caption to Fig. \ref{fig.Apparatus}. The linear and orthogonal polarization selection rules of the $\Ex$ and $\Ey$ transitions allow us to select the emission from one of these transitions in each NV center by inserting linear polarizers into the ZPL collection arms \cite{Fu09} and setting the Objective HWPs, shown in Fig. \ref{fig.Apparatus}, to the correct angles. Because the ZPL collection used to measure the HOM interference and the resonant excitation used to perform the PLE scans follow the same optical path, we can use the PLE scans to set the correct polarization angle for the ZPL collection. Therefore, we can selectively collect photons emitted from the desired transitions under non-resonant excitation with green light.

\begin{figure}
\begin{center}\includegraphics[width = \figSize \textwidth]{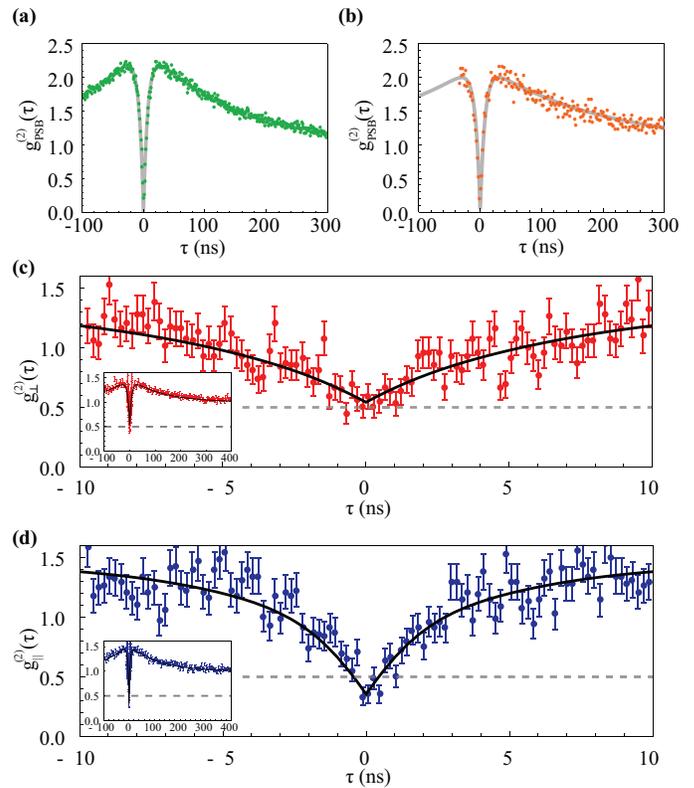}
\caption[G2sAndHOM]{(a,b) Single emitter second-order autocorrelation functions, $\gPSB(\tau)$, of the PSB emission, inferred for the two NV centers used for the HOM measurement. The FWHM of the central antibunching features are $7.5\pm0.1$ ns for (a) and $9.5\pm0.2$ ns for (b). (c,d) Demonstration of HOM interference from remote NV centers in the (c) distinguishable case and (d) indistinguishable case. The dashed lines indicate the limit expected from independent distinguishable single photon sources at $\tau=0$. Solid lines are a fit to the data based on the model described in the text. For the distinguishable case (c), we find that the FWHM of the central antibunching feature is $9.2\pm0.4$ ns and $\gPerp(0) = 0.54 \pm 0.04$. For the indistinguishable case (d), we find that the FWHM of the central antibunching feature is $5.6\pm0.3$ ns and $\gPar(0) = 0.35 \pm 0.04$.  Error bars are estimated based on shot noise. 
 }
\label{fig.G2sAndHOM}
\end{center}
\end{figure}

To confirm that we are addressing one single-photon emitter in each SIL, we infer the normalized, second-order autocorrelation function $\gPSB(\tau)$ in a standard Hanbury Brown and Twiss setup by splitting the PSB emission in a 50:50 beamsplitter. We expect $\gPSB(0)=0$ for an ideal single-photon source, and the single-photon nature of the emission is confirmed in Figs. \ref{fig.G2sAndHOM}(a,b). Resonant photons from each NV center is sent to an individual input port of a polarization-maintaining fiber-based beamsplitter. We balance the emission intensity by adjusting the green excitation intensity for each NV center independently to obtain 1100 counts per second (Cts/s) per emitter at each output port of the beamsplitter. An additional HWP in one setup is used to adjust the polarization matching of the photons at the beamsplitter. The output ports of the beamsplitter are connected to single photon detectors with timing resolution below 100 ps. The cross-correlation between these detectors is evaluated using a Time-Correlated Single Photon Counting Module with a resolution of 64 ps.

Ideally, $g^{(2)}(0)=0$ for a pair of indistinguishable photons, but the minimal observable $g^{(2)}(0)$ value increases in the presence of experimental noise, as described below. When the photons are distinguishable and the light intensity in both arms is balanced, the correlation measurement will yield $g^{(2)}(0) = 0.5$ \cite{Hong1987}. Measuring the cross-correlation function in the distinguishable case is equivalent to measuring the autocorrelation function of one emitter while the non-interfering emission from the other acts as uncorrelated noise, raising $g^{(2)}(0)$ from 0 to 0.5. Therefore, a measurement of $g^{(2)}(0) < 0.5$ indicates quantum interference between photons emitted by the two single photon sources.

Figures \ref{fig.G2sAndHOM}(c,d) show $g^{(2)}(\tau)$ for two different settings of the HWP angle. In Fig. \ref{fig.G2sAndHOM}(c), the angle is selected such that the emissions from the two NV centers are distinguished by their polarization, yielding $\gPerp(0) = 0.54 \pm 0.04$. In Fig. \ref{fig.G2sAndHOM}(d), the photons are indistinguishable when their polarizations are parallel, yielding $\gPar(0) = 0.35 \pm 0.04$. In terms of the visibility of the HOM interference, defined as $\eta=[\gPerp(0)-\gPar(0)]/\gPerp(0)$, we find $\eta=35\pm9\%$. This $\eta>0$ clearly demonstrates quantum interference between photons emitted by two NV centers separated by 2 m.


We next turn to the detailed discussion of our experimental observations.   We first consider the sources of noise that will cause our result to deviate from the ideal case $\gPar(0)=0$.  The APD dark counts and fluorescence background from our samples will lead to coincidence events, independent of the emission from the NV centers.  Background light and the dark counts of our detectors contribute 80 Cts/s out of the total 1100 Cts/s signal, raising $\gPar(0)$ to 0.14.  Because the NV center spin is not perfectly polarized under green illumination \cite{Manson2011}, we expect to collect emission from other transitions (e.g. $|A_2\rangle$ to $|m_s = \pm1\rangle$); since this emission is assumed to be circularly polarized, it is only partially filtered by the polarizer. Because our collection objective has a large numerical aperture, the polarizations of the $\Ex$ and $\Ey$ emissions are not perfectly orthogonal.  Using PLE spectra acquired during the HOM data acquisition period, we measure that the selected transitions contribute a minimum of 94\% of the total ZPL emission detected. Emission from other transitions at different frequencies raises the value of $\gPar(0)$ by 0.13. Finally the polarization-maintaining fiber-based beamsplitters introduce rotations to the polarization of the emission, which increases the distinguishability of the two photons. This contribution raises the $\gPar(0)$ value by 0.07.  Considering these factors, we expect experimental imperfections to raise the $\gPar(0)$ value to 0.34, which is in very good agreement with our experimental observations.

We now analyze the temporal behavior of the interference data presented in Figs. \ref{fig.G2sAndHOM}(c,d). The cross correlation function at the output ports of a beam splitter whose input ports are balanced and driven by single photon sources can be written as  \cite{Lettow2010}
\begin{eqnarray}
 & g^{(2)}(\tau) =  \frac{1}{4}\tilde{g}_{11}^{(2)}(\tau) +\frac{1}{4} \tilde{g}_{22}^{(2)}
(\tau)  \nonumber \\ + &\frac{1}{2}\left( 1 - \xi g_{11}^{(1)} (\tau)
 g_{22}^{(1)}(\tau) \cos \left( \Delta \omega \tau \right) \right),
\label{SandoghdarEqn}
\end{eqnarray}
where $\tilde{g}_{ii}^{(2)}(\tau)$ are the second-order autocorrelation functions for each single photon source inferred from Figs. \ref{fig.G2sAndHOM}(a,b) and $\Delta\omega$ is the frequency difference between the emitters. The slowly-varying component of the first-order correlation function for single photon sources is given by $g_{ii}^{(1)}(\tau) = \exp\left(-\gamma\left|\tau \right|/2  \right)$ where $\gamma\sim 1/12$ $\mathrm{ns}^{-1}$ is the inverse lifetime of the emitter. In our model, we assume that the emission from the two NV centers is radiatively broadened with bandwidth $\sim\gamma$, and that the center frequencies of the emitted photons are random and different for subsequent emissions.  We assume the distribution of the center frequencies is given by the Lorentzian profile that we fit to the PLE spectra shown in Fig. \ref{fig.Tuning}(d). Two features are visible in Fig. \ref{fig.G2sAndHOM}(d) for $\left|\tau \right|\leq10$ ns, which correspond to different terms in Eq. (\ref{SandoghdarEqn}). Quantum interference is described by the cosine term, which gives rise to a narrow interference feature, as described below.  This feature sits on top of a broader antibunching feature given by the second-order autocorrelation functions.  The amplitude of the interference term is fit using a phenomenological parameter $\xi$.  We convolve the quantum interference term with the frequency distribution, which washes out the $\cos \left( \Delta \omega \tau \right)$ oscillations and determines the $1/e$ full-width of the interference feature. We keep this width, which is determined independently by the PLE spectra to be 3.1 ns, constant for the fits in Figs. \ref{fig.G2sAndHOM}(c,d) and find excellent agreement. 

The behavior of the measured $\gPar(\tau)$ for $\left|\tau \right|  > 3.1/2$ ns is determined solely by $\tilde{g}_{ii}^{(2)}(\tau)$. Using the model described in \cite{Kurtsiefer2000}, we extract parameters, which are listed in the caption for Fig. \ref{fig.G2sAndHOM}, from the two-emitter cross-correlation datasets [Figs. \ref{fig.G2sAndHOM}(c,d)] that are in good agreement with those extracted from the single-emitter autocorrelation datasets [Figs. \ref{fig.G2sAndHOM}(a,b)]. The small difference between the parameters of Figs. \ref{fig.G2sAndHOM}(a,b) and Figs. \ref{fig.G2sAndHOM}(c,d) can be explained by drifts in laser intensity and focal spot position during the longer integration time used for the datasets shown in Figs. \ref{fig.G2sAndHOM}(c,d).

In summary, we have demonstrated the generation of indistinguishable photons from two spatially separated NV centers. Combined with the recent demonstration of entanglement between the electronic spin of an NV center and the polarization of a photon \cite{Togan2010}, our work paves the way for optically mediated generation of entanglement between remote solid-state quantum registers. The techniques demonstrated here have yielded improved collection efficiency, control of the NV centers' optical transition frequencies via electric field tuning, and the ability to operate two independent setups simultaneously over three days of continuous data acquisition. Additionally, implementing a resonant excitation scheme similar to that used in \cite{Togan2010} will likely result in narrow optical linewidths \cite{Fu09}.  Such a scheme will also minimize the ionization of local charge traps, which increases measurement time while using electric field tuning with CW excitation at 532 nm. The important figure of merit for an entanglement experiment is the time required to generate an entangled pair with fidelity greater than 50\%. Using our currently available values for collection efficiency ($4\times10^{-5}$), narrow linewidths ($50$ MHz), and assuming a repetition rate of $10^8$, we estimate that one entangled spin pair can be created within roughly ten seconds. Improved photon collection techniques that are currently being developed \cite{Aharonovich2011} have the potential to increase this generation rate dramatically.  Even with the currently estimated rates, though, the exceptionally long nuclear spin memory times of NV centers \cite{Maurer} may allow one to use such systems for the realization of solid-state, multi-node quantum networks.


We thank N. de Leon, J. D. Thompson, R. Hanson, and L. Childress for useful discussions. This work was supported by NSF, CUA, DARPA QUEST program, AFOSR MURI, and Packard Foundation. A.K. acknowledges support from the Alexander von Humboldt Foundation. The electrodes for electric field tuning were fabricated in the Center for Nanoscale Systems (CNS) at Harvard University.

\end{document}